\documentclass[usenatbib]{emulateapj}
\usepackage{graphicx}
\usepackage[flushleft]{threeparttable}
\usepackage[usenames,dvipsnames,svgnames,table]{xcolor}
\usepackage{hyperref}
\definecolor{darkblue}{rgb}{0.0,0.0,0.3}
\hypersetup{colorlinks,breaklinks,
            linkcolor=darkblue,urlcolor=darkblue,
            anchorcolor=darkblue,citecolor=darkblue}
\usepackage{amsmath,amssymb}
\usepackage{natbib}

\newcommand{\pD}[2]{\frac{\partial #2}{\partial #1}}

\newcommand\bb[1]{\mbox{\boldmath{$#1$}}}
\newcommand\grad{\bb{\nabla}}
\newcommand\bcdot{\,\bb{\cdot}\,}
\newcommand\btimes{\,\bb{\times}\,}
\newcommand{\mc}[1]{\mathcal{#1}}

\newcommand\bs[1]{\boldsymbol{#1}}

\newcommand{\ez}{\hat{\bb{z}}}

\newcommand{\rmd}{{\rm d}}

\newcommand{\rhoi}{\rho_{{\rm i}0}}
\newcommand{\valf}{v_{\rm A}}
\newcommand{\valfo}{v_{{\rm A}0}}
\newcommand{\di}{d_{{\rm i}0}}

\newcommand{\vthio}{v_{{\rm thi}0}}
\newcommand{\betaio}{\beta_{\rm i0}}

\begin{document}

\newcommand\cmtmk[1]{{\color{red}[MK: #1]}}
\newcommand\cmtla[1]{{\color{blue}[LA: #1]}}


\title{Hybrid-Kinetic Simulations of Ion Heating in Alfv\'{e}nic Turbulence}

\author{Lev Arzamasskiy\altaffilmark{1,*}, Matthew W.~Kunz\altaffilmark{1,2}, Benjamin D.~G.~Chandran\altaffilmark{3}, and Eliot Quataert\altaffilmark{4}}
\altaffiltext{1}{Department of Astrophysical Sciences, 
Princeton University, Ivy Lane, Princeton, NJ 08540}
\altaffiltext{2}{Princeton Plasma Physics Laboratory, P.O.~Box 451, Princeton, NJ 08543, USA}
\altaffiltext{3}{Department of Physics, University of New Hampshire, 242B Morse Hall, 8 College Rd, Durham, NH 03824, USA}
\altaffiltext{4}{Department of Astronomy \& Theoretical Astrophysics Center, University of California, 501 Campbell Hall \#3411, Berkeley, CA 94720-3411, USA}
\altaffiltext{*}{\email{leva@astro.princeton.edu}}


\begin{abstract}
We present three-dimensional, hybrid-kinetic numerical simulations of driven Alfv\'{e}n-wave turbulence of relevance to the collisionless near-Earth solar wind. Special attention is paid to the spectral transition that occurs near the ion-Larmor scale and to the origins of preferential perpendicular ion heating and of non-thermal wings in the parallel distribution function. Several novel diagnostics are used to show that the ion heating rate increases as the kinetic-Alfv\'{e}n-wave fluctuations, which comprise the majority of the sub-ion-Larmor turbulent cascade, attain near-ion-cyclotron frequencies. We find that ${\approx}75$--$80\%$ of the cascade energy goes into heating the ions, broadly consistent with the near-Earth solar wind. This heating is accompanied by clear velocity-space signatures in the particle energization rates and the distribution functions, including a flattened core in the perpendicular-velocity distribution and non-Maxwellian wings in the parallel-velocity distribution. The latter are attributed to transit-time damping and the pitch-angle scattering of perpendicularly heated particles into the parallel direction. Accompanying these features is a steepening of the spectral index of sub-ion-Larmor magnetic-field fluctuations beyond the canonical $-2.8$, as field energy is transferred to thermal energy. These predictions may be tested by measurements in the near-Earth solar wind.
\end{abstract}




\section{Introduction.} 
\label{sect:intro}
\setcounter{footnote}{0}
It has been 48 years since NASA's {\it Mariner 5} established definitively that the interplanetary medium plays host to a broadband spectrum of large-amplitude Alfv\'{e}n waves propagating outwards from the Sun (\citealt{bd71}, following pioneering work using {\it Mariner 2} data by \citealt{coleman68}). We now know that the solar wind is turbulent \citep{tm95,goldstein95}, exhibiting a power spectrum extending over several decades in scale \citep{bc05,alexandrova13}. Most of the energy at large scales is in the form of Alfv\'{e}nic fluctuations, which have magnetic and velocity fields perpendicular to the mean magnetic-field direction. As this energy cascades down to smaller scales, an inertial range is set up, one whose defining characteristic is anisotropy with respect to the field direction \citep{matthaeus90,bieber96,horbury08,podesta09,wicks10,chen11a}. This anisotropy, central to modern theories of magnetohydrodynamic (MHD) turbulence \citep{zank1992,zank1993,gs95,boldyrev06,chandran15,ms17} and manifest in the observed shapes of turbulent eddies and their spectral slopes \citep{horbury12}, extends all the way down to kinetic scales \citep{sahraoui10,chen10}, where the turbulence is ultimately dissipated as heat.

The amplitudes of turbulent fluctuations measured in the solar wind are positively correlated with solar-wind temperature \citep{grappin90} and imply a turbulent heating rate comparable to the observationally inferred solar-wind heating rate \citep{smith01,breech09,cranmer09,stawarz09}. While these observations establish a close connection between the global evolution of the solar wind and the dissipation of turbulence within it, the precise nature of this dissipation is puzzling. Minor ions in coronal holes and protons in low-beta fast-solar-wind streams are heated in such a way that thermal motions perpendicular to the background magnetic field are more rapid than thermal motions along it (i.e., $T_\perp > T_\parallel$; \citealt{kohl98,li98,antonucci00,marsch82,marsch04,hellinger06}). Moreover, the evolution of proton temperature anisotropy from $0.3~{\rm au}$ to $0.9~{\rm au}$ clearly indicates non-adiabatic particle heating preferentially in the field-perpendicular direction (see Fig.~1 of \citealt{matteini07}).

These observations present a challenge for models of solar-wind heating based upon theories of Alfv\'{e}nic turbulence, which predict an anisotropic cascade of energy primarily to small perpendicular (rather than parallel) scales (i.e., $k_\perp \gg k_\parallel$; \citealt{shebalin83,gs95,nb96,matthaeus98,galtier00,cho02,mg01}). Such an anisotropic cascade is inefficient at transporting energy to high frequencies traditionally considered necessary to explain the observed strong perpendicular heating (e.g., via ion-cyclotron resonant heating; \citealt{quataert1998,leamon98a,isenberg01,mt01,hi02,kasper13,cranmer14}).

An alternative explanation for the measured strong perpendicular heating of ions is that of low-frequency stochastic heating, which arises when particles interact with turbulent fluctuations whose characteristic frequencies are much smaller than the cyclotron frequency but whose amplitudes at the Larmor scale (i.e., $k_\perp \rho_{\rm i} \sim 1$ for ions) are sufficiently large \citep{mcchesney87}. The particle's motion in the field-perpendicular plane then becomes chaotic instead of quasi-periodic \citep{kruskal62}, leading to diffusion in the perpendicular energy space due to interactions with the time-varying electrostatic potential \citep{jc01,chen01,white02,vg04,bourouaine08,chandran10}. In the context of solar-wind turbulence, low-frequency stochastic heating has been studied numerically using test particles in randomly phased kinetic Alfv\'{e}n waves (KAWs) \citep{chandran10} and in reduced-MHD turbulence \citep{xia13}, and observationally in {\em Helios-2} and {\em Wind} data \citep{chandran13,bc13,vech17}. It was also the focus of work by \citet{vasquez15}, who used hybrid-kinetic simulations of decaying turbulence in a low-beta, cold-electron plasma to find that the perpendicular heating rate scales with the cube of the turbulence amplitude evaluated at the ion-Larmor scale (as predicted by \citealt{chandran10}, but without their multiplicative factor that exponentially suppresses stochastic heating at small enough turbulence amplitude).

In this paper, we analyze ion heating in three-dimensional (3D) hybrid-kinetic simulations of driven, quasi-steady-state, magnetized turbulence, resolving scales above and below the ion Larmor radius. This analysis is done self-consistently, in that the evolution of the distribution function due to the ions' interactions with the electromagnetic fields in turn modifies those fields in a reciprocal fashion (as opposed to test-particle calculations, in which the particles' evolution does not feed back on the fluctuations driving it). Our goal is to understand the relationship between the properties of kinetic, Alfv\'{e}nic turbulence and the character of the ion heating occurring within it. For this, it is important to note that our simulations do not adopt the oft-employed gyrokinetic approximation \citep[see, e.g.,][]{howes06,schekochihin09}, and so we allow for electromagnetic fluctuations having finite amplitudes on ion-Larmor scales and/or having ion-Larmor frequencies. This is crucial, as it is quantitatively unclear to what extent the observed anisotropy of solar-wind turbulence precludes an appreciably energetic component of high-frequency electromagnetic fluctuations \citep[e.g.,][]{he11}.

An outline is as follows. In \S \ref{sect:hybrid} we detail our numerical method and state the parameters used in the runs. \S \ref{sect:results} catalogues our results, whose interpretation is taken up in \S \ref{sect:interpretation}. We close in \S \ref{sect:summary} with a summary of our results, placed in the context of particle heating in the turbulent solar wind and other hot, dilute astrophysical plasmas.


\section{Hybrid-kinetic Simulations of Driven Alfv\'{e}nic Turbulence}
\label{sect:hybrid}

We consider a non-relativistic, quasi-neutral, collisionless, and initially homogeneous plasma of kinetic ions (mass $m_{\rm i}$, charge $e$) and massless fluid electrons threaded by a uniform magnetic field $\bb{B}_0 = B_0 \ez$ and subject to a stochastic driving force $\bb{F}(t,\bb{r})$. The model equations governing the evolution of the ion distribution function $f_{\rm i}(t, \bb{r}, \bb{v})$ and the magnetic field $\bb{B}(t,\bb{r})$ are, respectively, the Vlasov equation,
\begin{equation}\label{eqn:vlasov}
\pD{t}{f_{\rm i}} + \bb{v} \bcdot \grad f_{\rm i} + \left[ \frac{e}{m_{\rm i}} \left( \bb{E} + \frac{\bb{v}}{c} \btimes \bb{B} \right) + \frac{\bb{F}}{m_{\rm i}} \right]\! \bcdot \pD{\bb{v}}{f_{\rm i}} = 0 ,
\end{equation}
and Faraday's law of induction,
\begin{equation}\label{eqn:induction}
\pD{t}{\bb{B}} = - c \grad \btimes \bb{E} .
\end{equation}
The electric field,
\begin{equation}\label{eqn:efield}
\bb{E} = - \frac{\bb{u}_{\rm i} \btimes \bb{B}}{c} + \frac{( \grad \btimes \bb{B} ) \btimes \bb{B}}{4\pi e n_{\rm i}} - \frac{T_{\rm e} \grad n_{\rm i}}{e n_{\rm i}},
\end{equation}
is obtained by expanding the electron momentum equation in $( m_{\rm e} / m_{\rm i} )^{1/2}$, enforcing quasi-neutrality, 
\begin{equation}\label{eqn:quasineutrality}
n_{\rm e} = n_{\rm i} \equiv \! \int {\rm d}^3 \bb{v} \, f_{\rm i} ,
\end{equation}
assuming isothermal electrons ($T_{\rm e} = {\rm const}$), and using Amp\'{e}re's law to solve for the mean electron velocity
\begin{equation}\label{eqn:ampere}
\bb{u}_{\rm e} = \bb{u}_{\rm i} - \frac{\bb{j}}{e n_{\rm i}} \equiv  \frac{1}{n_{\rm i}} \int {\rm d}^3 \bb{v} \, \bb{v} f_{\rm i} - \frac{ c \grad \btimes \bb{B} }{4 \pi e n_{\rm i}}
\end{equation}
in terms of the mean ion velocity $\bb{u}_{\rm i}$ and the current density $\bb{j}$. Equations (\ref{eqn:vlasov})--(\ref{eqn:ampere}) constitute the hybrid-kinetic model of Vlasov ions and (massless) fluid electrons \citep{byers78,hn78}. 

These equations are solved using the second-order--accurate, particle-in-cell code {\sc Pegasus} \citep{kunz14}. A nonlinear $\delta f$ method is used to reduce the impact of finite-particle-number noise on computed moments of $f_{\rm i}$. To remove small-scale power from the fluctuating fields, the zeroth ($n_{\rm i}$) and first ($n_{\rm i}\bb{u}_{\rm i}$) moments of $f_{\rm i}$ are low-pass filtered once per time step. A fourth-order hyper-resistivity is included to remove grid-scale magnetic energy, its value tuned to achieve quasi-steady state.

Observations of turbulence in the solar wind  show that it consists of a majority of nearly incompressible, Alfv\'{e}nically polarized, spatially anisotropic fluctuations \citep[typical density variations $ \lesssim 10\%$;][]{celnikier1983,roberts1987,mt1990,tm1994,bieber96}. To excite such fluctuations and minimize the excitation of compressible motions, $\bb{F}$ is oriented in the $x$-$y$ plane perpendicular (``$\perp$'') to $\bb{B}_0$ 
and constrained to satisfy $\grad\bcdot\bb{F} = 0$. At each simulation time step, the Fourier coefficients $\bb{F}_{\bs{k}}$ are independently generated from a Gaussian-random field with power spectrum $k^{-5/3}_\perp$ in the wavenumber range $k_z \in [1,2] \times \left(2\pi / L_z\right)$ and $k_{(x,y)} \in [1,2] \times (2\pi/L_{(x,y)})$, where $L_{(x,y,z)}$ is the size of the periodic computational domain in the ($x,y,z$) direction. The resulting force is then inverse-Fourier transformed, shifted to ensure no net momentum injection, and normalized to provide power per unit volume $\dot{\varepsilon}$. The force is time-correlated over $t_{\rm corr}$ using an Ornstein--Uhlenbeck process: 
\begin{equation}
\bb{F}(t+\rmd t,\bb{r}) = \theta \bb{F}(t,\bb{r}) + \sqrt{1-\theta^2} \bb{F}'(\bb{r}) ,
\end{equation}
which has the autocorrelation (in the limit $\rmd t\rightarrow 0$)
\begin{equation}
\langle \bb{F}(t_1,\bb{r}) \bcdot \bb{F}(t_2,\bb{r}) \rangle = \langle |  \bb{F}(t,\bb{r}) |^2 \rangle \, e^{(t_1-t_2)/t_{\rm corr}} ,
\end{equation}
where $\rmd t$ is the timestep, $\theta \equiv \exp( - \rmd t / t_{\rm corr} )$,  $t_{\rm corr}$ is the correlation time of the driving, and $\bb{F}'$ is a new Gaussian-random field generated as detailed above. Time-correlated driving avoids spurious particle acceleration via resonances with high-frequency power in, e.g., $\delta$-correlated driving \citep{lynn12}.

Most of the power in strong MHD turbulence resides in fluctuations satisfying $k_\parallel / k_\perp \lesssim u_\perp(k_\perp) / \valf$ \citep[``critical balance'';][]{gs95,mallet15}, where $k_\parallel$ ($k_\perp$) is the wavenumber parallel (perpendicular) to the local magnetic field and $\valf \equiv B / (4\pi m_{\rm i} n_{\rm i})^{1/2}$ is the Alfv\'{e}n speed. We therefore choose $\dot{\varepsilon}$ and $L_{(x,y,z)}$ such that, in saturation, the rms velocity fluctuation $u_{\rm rms}$ satisfies $u_{\rm rms} / \valf \simeq L_x / L_z = L_y / L_z \ll 1$ at the outer scale; likewise, $t_{\rm corr} = L_z/2\pi\valf \simeq L_x/2\pi u_{\rm rms}$. This is meant to mimic an energy cascade from larger scales that is present in the solar wind, whose inertial-range turbulence is consistent with critical balance \citep{horbury08,podesta09,lw10,wicks10}. 

By replacing electron kinetics with an isothermal equation of state, the hybrid-kinetic model excludes electron Landau damping and its effect on the turbulent cascade \citep[e.g.,][]{tenbarge13,told16a,grovselj17}. However, the hybrid-kinetic model affords a huge cost savings over the fully kinetic approach (see \citet{vasquez14}, \citet{cerri17}, and \citet{franci18} for recent examples of using the hybrid-kinetic approach to simulate 3D solar-wind-like turbulence, and \citet{parashar09} for an early hybrid-kinetic approach to studying particle heating in a 2D Orszag--Tang vortex). And it captures physics not described by the oft-employed gyrokinetic approach \citep[e.g.,][]{howes06,howes08,howes11,schekochihin09,told15,kawazura18}, such as stochastic ion heating, ion-cyclotron resonances, and modes whose propagation angles are not asymptotically oblique. We refer the reader to \citet{told16b} and \citet{cb17} for comparison of linear modes in hybrid-kinetics, gyrokinetics, and full kinetics.

We present results from two simulations: $\betaio\equiv 8\pi n_{{\rm i}0} T_{{\rm i}0} / B^2_0 = 0.3$ and $1$, both with $T_{\rm e} = T_{{\rm i}0}$ (the subscript ``0'' denotes an initial value). For $\betaio = 1$, $N_{\rm ppc} = 512$ particles per cell were drawn from a Maxwell distribution and placed on a 3D periodic grid of $N_x \times N_y \times N_z = 200^2 \times 1600$ cells spanning $L_x \times L_y \times L_z = (20\pi\rhoi)^2 \times 160\pi\rhoi$, where $\vthio \equiv (2T_{{\rm i}0}/m_{\rm i})^{1/2}$ is the ion thermal speed, $\rhoi \equiv \vthio / \Omega_{{\rm i}0}$ is the ion Larmor radius, and $\Omega_{{\rm i}0} \equiv eB_0 / m_{\rm i} c$ is the ion gyrofrequency. These parameters provide reasonable scale separation between the grid scale, the ion-kinetic scales, and the driving scales: $k_z \di \in [0.0125,10]$ and $k_{(x,y)} \di \in [0.1,10]$, where $\di = \rhoi / \betaio^{1/2}$ is the ion skin depth. For $\betaio = 0.3$, $N_{\rm ppc} = 216$, $N_x \times N_y \times N_z = 200^2 \times 1200$, and $L_x \times L_y \times L_z = (20\pi\rhoi)^2 \times 120\pi\rhoi$. These imply $k_z \di \in [0.03,18.26]$ and $k_{(x,y)} \di \in [0.18,18.26]$.

With these scales borne in mind, it is useful to compare with the observed spectral anisotropy in the near-Earth solar wind. There, spectral anisotropy of the turbulent cascade begins around $k\rho_{\rm i} \sim 10^{-3}$ and increases as $k_\parallel/k_\perp \propto k^{-1/3}_\perp$ towards smaller scales \citep{wicks10}. Adopting these scalings implies $k_\parallel/k_\perp \simeq 0.17$ around $k_\perp\rho_{\rm i} \simeq 0.2$, where our simulated inertial range begins. Thus, our chosen box aspect ratios of $1/8$ (for $\betaio=1$) and $1/6$ (for $\betaio=0.3$) accurately capture the spectral anisotropy of solar-wind turbulence arriving from larger scales to near ion-Larmor scales. (This point is revisited near the end of \S\ref{sect:fourier} in the context of the observed spectral anisotropy {\em at} ion-Larmor scales.)

Both simulations required at least $2L_z/\valfo \equiv 2t_{\rm cross}$ to obtain a quasi-steady state, in which the properties of the turbulence exhibit only minimal secular evolution. An additional ${\sim}10t_{\rm cross}$ were run to procure statistically converged results ({\it viz.}, $10t_{\rm cross}$ for $\betaio=1$ and $18t_{\rm cross}$ for $\betaio=0.3$). Note that the particle distribution function never reaches a true steady state, as there is no cooling and so the total particle energy grows monotonically in time. However, the changes in ion temperature at the end of both simulations are small enough to affect neither the properties of the turbulence nor the heating diagnostics measured in these simulations. In what follows, $\langle\,\cdot\,\rangle$ denotes a spatio-temporal average over all cells measured in this quasi-steady state.


\section{Results}  
\label{sect:results}

%
%
\begin{figure}
    \centering
    \includegraphics[width=0.45\textwidth]{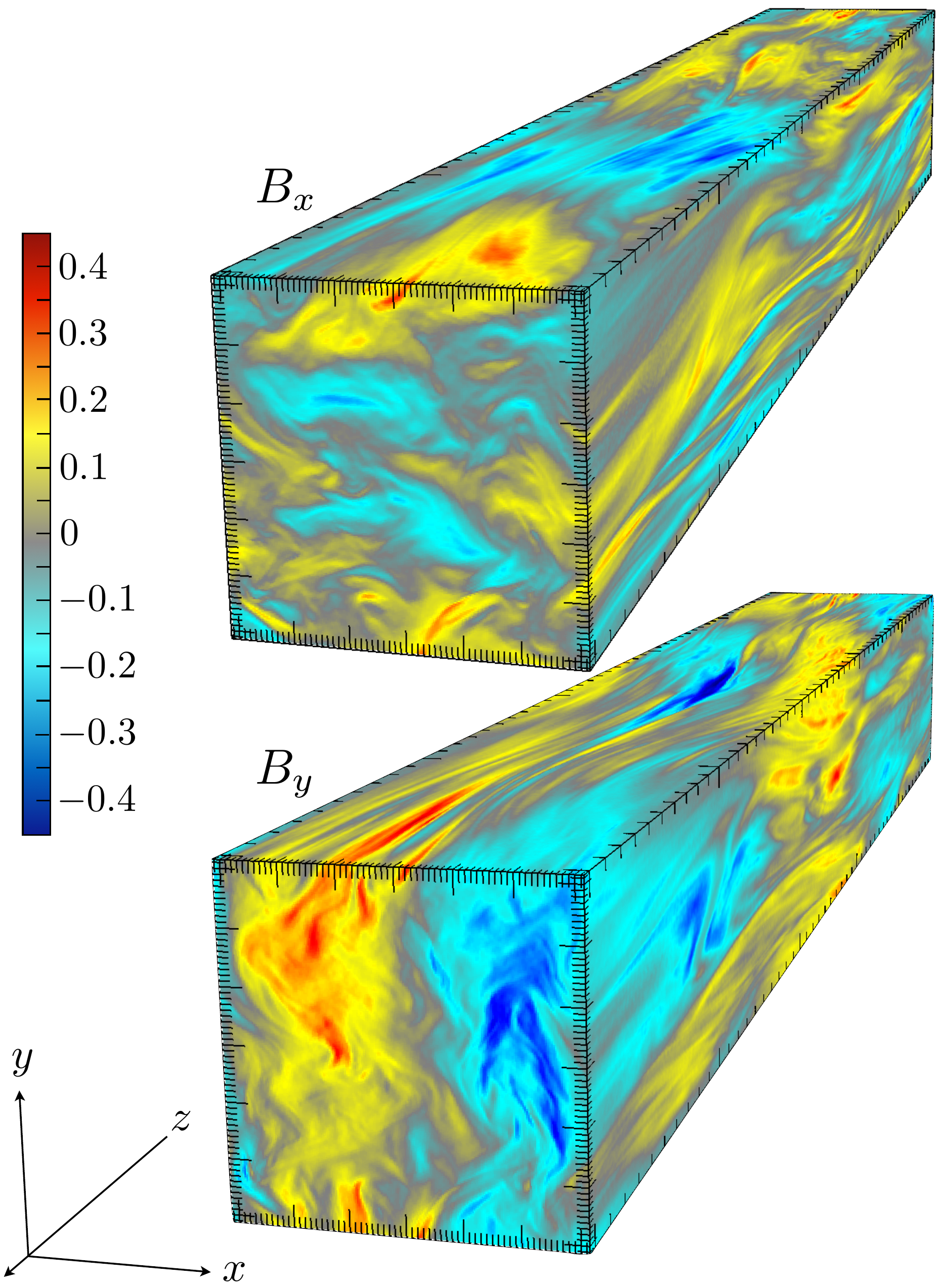}
    \caption{Pseudo-color images of the $x$- and $y$-components of the fluctuating magnetic field perpendicular to the guide field, taken in quasi-steady state. Field strengths are normalized to $B_0$.}
    \label{fig:images}
\end{figure}

%
%
\begin{figure*}
\centering
\includegraphics[width=0.95\textwidth]{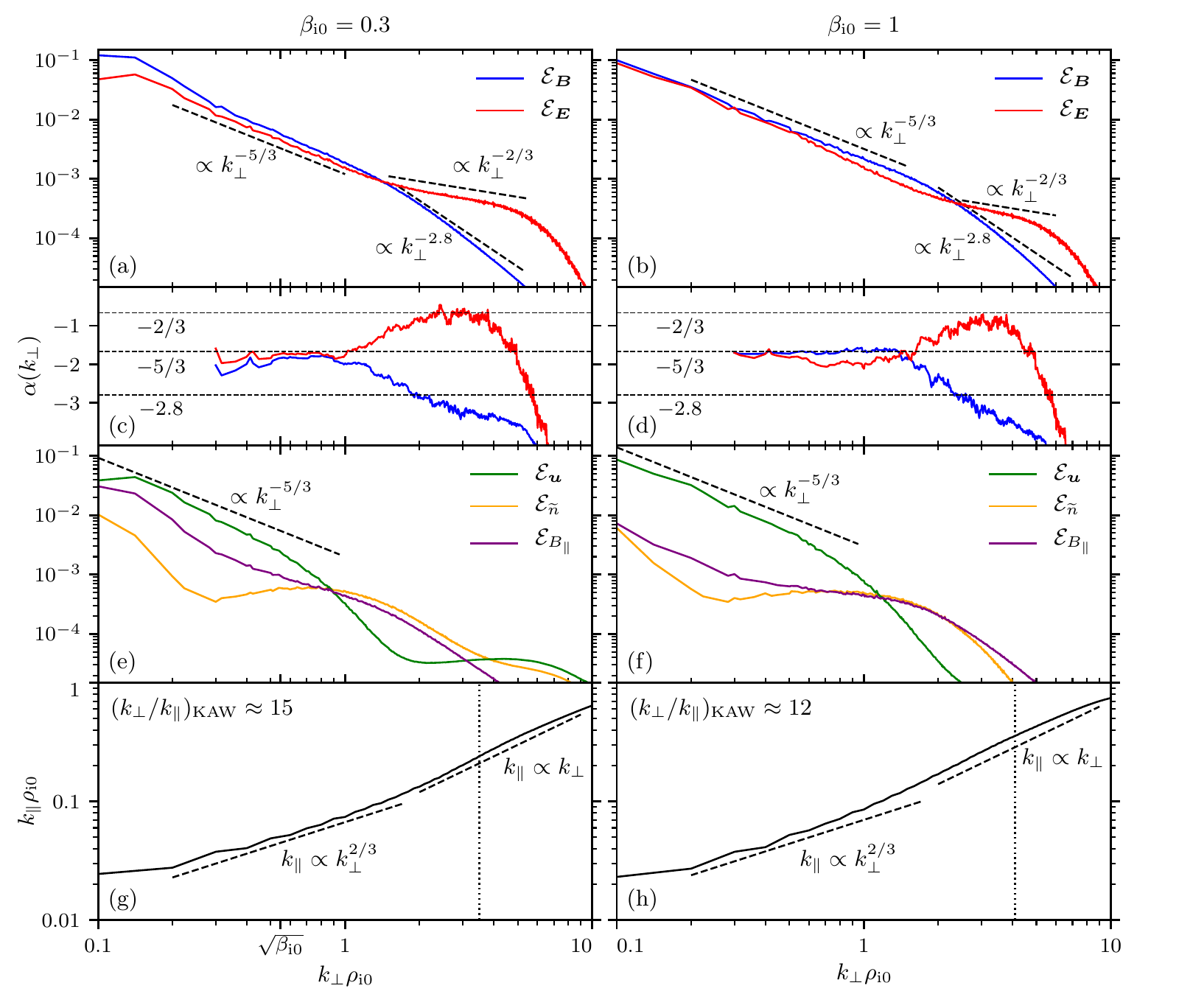}
\caption{(a,b) Energy spectra of magnetic (blue) and electric (red) fields versus $k_\perp$ for $\betaio = 0.3$ and $\betaio = 1$; their spectral indices $\alpha(k_\perp)$ are shown in panels (c) and (d). Reference slopes are provided. (e,f) Energy spectra of ion flow velocity (orange), normalized density $\widetilde n \equiv n_{\rm i}\sqrt{\betaio(1+\betaio)}$ (green), and parallel-magnetic-field fluctuations (purple). Here, perpendicular ($\perp$) and parallel ($\parallel$) are measured with respect to $\bb{B}_0$. (g,h) Spectral anisotropy of the magnetic-field fluctuations with respect to the scale-dependent {\em local} mean magnetic field, $k_\parallel(k_\perp)$, computed using the method devised in \citet{cho09}. Vertical dotted lines mark the values of $k_\perp$ at which $\omega_{\rm KAW} = k_\parallel v_{\rm A} k_\perp \rho_{\rm i} / \sqrt{1+\beta_{\rm i}} \approx \Omega_{\rm i0}$.
\label{fig:spectra}}
\end{figure*}

An example of the turbulent quasi-steady state is given in Figure \ref{fig:images}, which shows pseudo-color images of the $x$- and $y$-components (i.e., those perpendicular to the mean field) of the fluctuating magnetic field from the $\betaio=1$ run. Spatial anisotropy is evident, with short perpendicular scales and long parallel scales consistent with a critically balanced cascade (i.e., $\delta B_{\rm rms}/B_0 \approx 1/8$, the box aspect ratio). This anisotropy plays an important role in shaping the energy spectra (\S\ref{sect:fourier}) and the nature of ion heating (\S\ref{sect:heating}).

\subsection{Energy spectra and spectral anisotropy}
\label{sect:fourier}

%
%
\begin{figure}
\centering
\includegraphics[width=0.45\textwidth]{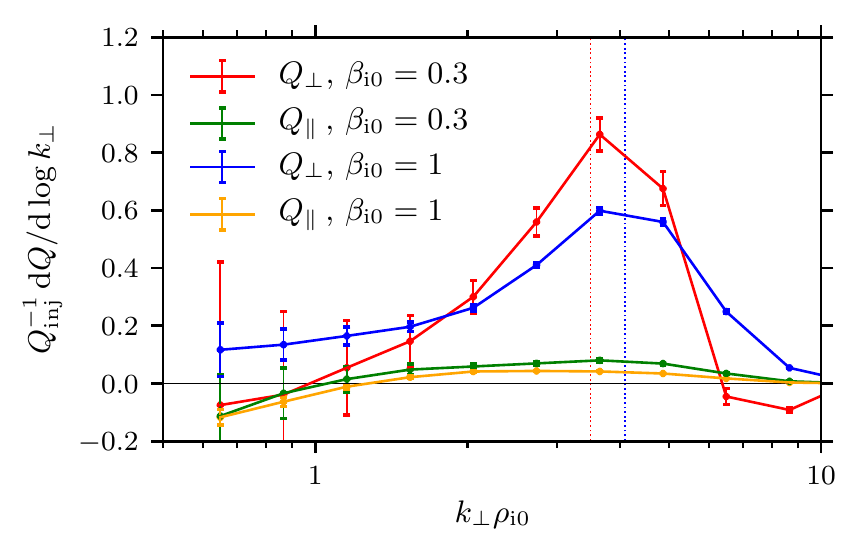}
\caption{Perpendicular ($Q_\perp$) and parallel ($Q_\parallel$) particle-energization rates as functions of $k_\perp\rhoi$ for both runs. (Here, ``$\perp$" and ``$\parallel$" are measured with respect to the guide field). Rates are measured within logarithmic $k_\perp$-bins centered on the points shown in the plot and are normalized to the energy injected by external driving, $Q_{\rm inj}$. For both $\betaio$, $Q_\parallel \ll Q_\perp$, with the latter peaking at $k_\perp\rhoi\approx 4$--$5$, near where $\omega_{\rm KAW} \approx \Omega_{\rm i0}$ in both runs (see \S\ref{sect:heating}). Error bars indicate the variance of $Q$ over time. As in Figure \ref{fig:spectra}(g,h), vertical dotted lines mark the values of $k_\perp$ at which $\omega_{\rm KAW} \approx \Omega_{\rm i0}$.
\label{fig:heating_vs_k}
}
\end{figure}

Figures \ref{fig:spectra}(a,b) present energy spectra of magnetic-field ($\mc{E}_{B}$) and electric-field ($\mc{E}_{E}$) fluctuations for $\betaio = 0.3$ and $\beta_{\rm i0} = 1$ versus the wavenumber $k_\perp$ perpendicular to $\bb{B}_0$. Their scale-dependent spectral indices $\alpha(k_\perp)$, computed about each $k_\perp$ using 21 neighboring points (except for the first 5 points for which we use 11 neighboring harmonics), are shown in Figures \ref{fig:spectra}(c,d). Accompanying these in Figures \ref{fig:spectra}(e,f) are the ion-flow-velocity ($\mc{E}_{u}$), density ($\mc{E}_{\widetilde{n}}$), and parallel-magnetic-field ($\mc{E}_{B_\parallel}$) energy spectra. 

In the inertial range ($k_\perp\rhoi\lesssim 1$), the spectral slopes are close to $-5/3$, the spectral slope predicted for an anisotropic, critically balanced cascade of Alfv\'{e}n waves \citep{gs95}. A spectral break occurs at $k_\perp\rhoi\sim 1$, at which point $\mc{E}_E$ flattens to take on a slope near $-2/3$, in agreement with measurements that show a flattening of the solar-wind electric-field power spectrum in ion-kinetic range \citep{bale05,sahraoui09,salem12} and predictions for intermittent KAW turbulence \citep{bp12}. In this range, $\mc{E}_B$ also steepens to take on a slope comparable to the $-2.8$ commonly observed in the near-Earth solar wind  \citep[e.g.,][]{sahraoui09,alexandrova09,alexandrova12} and within the range $[-2.5,-3.1]$ found in {\em Cluster} spacecraft measurements of the $\beta \sim 1$ solar wind \citep{sahraoui13}.
 However, this spectral steepening appears to be $k_\perp$-dependent, an attribute we revisit below when discussing ion heating. For now, we remark that such steepening is consistent with the sub-ion spectrum measured in Earth's magnetosheath \citep{chen2019}. Despite $-2.8$ being steeper than predictions for standard ($-7/3$; \citealt{schekochihin09}) and intermittent ($-8/3$; \citealt{bp12}) KAW turbulence, the normalized perturbed density $\delta\widetilde{n} \equiv ( \delta n_{\rm i} / n_{\rm i0} ) \sqrt{\betaio(1+\betaio)}$ approximately follows the linear KAW eigenfunction $\delta\widetilde{n} = \delta B_\perp/B_0 = (\delta B_\parallel/B_0) \sqrt{1+1/\betaio}$ (for $T_{\rm e0} = T_{\rm i0}$; \citealt{schekochihin09,bp13}). This suggests that the sub-ion-Larmor-scale cascade is primarily composed of KAWs (at least up to $k_\perp \rhoi \approx 3$ where $\mc{E}_{\widetilde{n}}$ starts to be affected by particle noise and digital filters), in agreement with combined analyses of magnetic-field fluctuations and of electric-field or density fluctuations in the ion-kinetic range of solar-wind turbulence \citep[e.g.,][]{salem12,chen13}. That being said, there is slightly more (about a factor ${\lesssim}2$) magnetic-fluctuation energy than anticipated for KAWs, suggesting the presence of additional wave modes (an excess of ${\simeq}1.33$ was measured in the solar wind by \citet{chen13}). At $k_\perp\rhoi\approx 6$, the spectra steepen further due to hyper-resistivity and spectral filters (ion heating is less important at these scales---see \S\ref{sect:heating}).

The predicted slopes of $-5/3$ in the inertial range and $-7/3$ (or $-8/3$) in the KAW range follow from assuming locality of interactions and constant energy flux in Fourier space, combined with a model for the spatial anisotropy of the turbulent fluctuations, $k_\parallel(k_\perp)$. The latter is afforded by the critical balance assumption, which states that the scale-dependent nonlinear cascade time is comparable to the linear timescale of the dominant wave at that scale \citep{gs95}---essentially a causality argument \citep{boldyrev05}. For the inertial range, in which the characteristic linear frequency is that of Alfv\'{e}n waves, $\omega_{\rm AW} = k_\parallel v_{\rm A}$, a perpendicular spectrum of $-5/3$ corresponds to $k_\parallel \propto k^{2/3}_\perp$. For the KAW range, in which the characteristic linear frequency is $\omega_{\rm KAW} = k_\parallel \valf k_\perp \rho_{\rm i}/\sqrt{1+\beta_{\rm i}}$, a perpendicular spectrum of $-7/3$ ($-8/3$) corresponds to $k_\parallel\propto k^{1/3}_\perp$ ($k_\parallel\propto k^{2/3}_\perp$), the steeper parenthetical values reflecting the  two-dimensionalization of intermittent KAW turbulence \citep{bp12}. 

In Figure \ref{fig:spectra}(g,h), we show the spectral anisotropy computed from our simulations.\footnote{To compute the spectral anisotropy, we use equation (34) of \citet{cho09} to compute the scale-dependent wavenumber parallel to the local mean magnetic field, $k_\parallel(k_\perp)$; those authors showed that this method works very well for steep spectra. We also computed two- and three-point second-order structure functions and measured the spectral anisotropy from their isocontours (similar to \citealt{cv00}), finding similar results.} In both runs, $k_\parallel \propto k^{2/3}_\perp$ in the inertial range. At the start of the KAW range, the wavevector anisotropy  $(k_\perp/k_\parallel)_{\rm KAW} \approx 15$ and $12$ for $\betaio = 0.3$ and $1$, respectively, corresponding to  wavevector obliquities $\theta_{(\bs{k},\bs{B})} \equiv \tan^{-1}(k_\perp/k_\parallel) \approx 86^\circ$ and $85^\circ$. These values are comparable to those measured in the solar wind \citep[e.g.,][]{sahraoui10,narita11}. Well into the sub-ion-Larmor range, however, we observe $k_\parallel \propto k_\perp$, steeper than current theoretical predictions and corresponding to a scale-independent anisotropy. Constant spectral anisotropy in the sub-ion-Larmor-scale range has also been seen in other hybrid-kinetic simulations of 3D Alfv\'{e}nic turbulence \citep{franci18}.

Constant spectral anisotropy at $k_\perp\rhoi > 1$ implies that the turbulence there is more likely to attain ion-Larmor frequencies before reaching electron scales, since $\omega_{\rm KAW} \propto k^{2}_\perp$, rather than the standard $\omega_{\rm KAW} \propto k^{4/3}_\perp$ predicted by theories of low-frequency gyrokinetic turbulence \citep[e.g.,][]{schekochihin09}. Such high-frequency fluctuations facilitate additional energy-transfer channels that are not present in gyrokinetics \citep[cf.][]{howes08heating}, such as cyclotron resonances and high-frequency stochastic heating, to which we now turn.

\subsection{Ion heating}
\label{sect:heating}

%
%
\begin{figure*}
\centering
~~\includegraphics[width=0.95\textwidth]{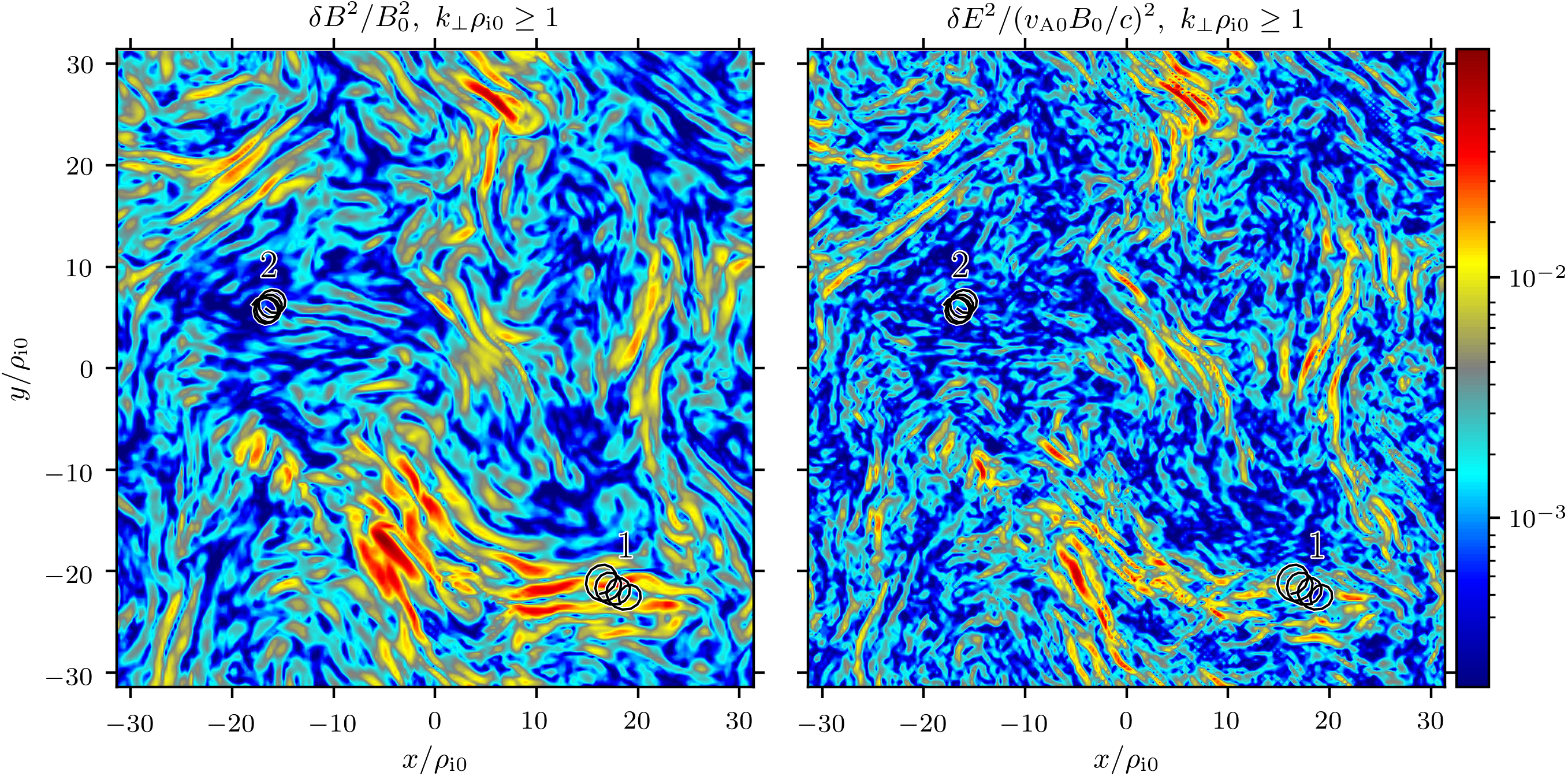}
\includegraphics[width=0.95\textwidth]{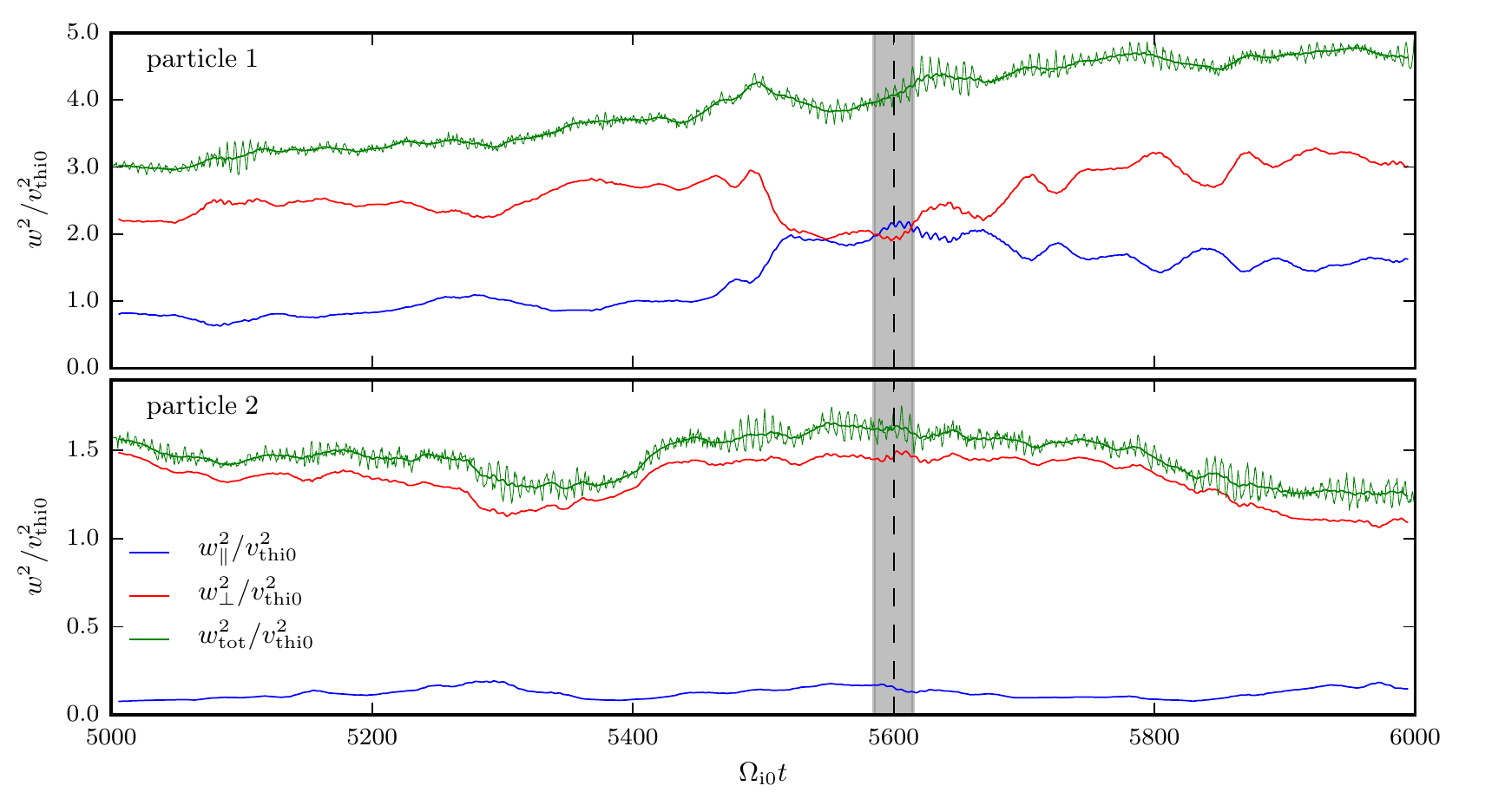}
\caption{Example particle orbits from the $\betaio = 1$ run. (top) Snapshots of magnetic-field (left) and electric-field (right) fluctuations in a plane perpendicular to the guide field on scales $k_\perp \rhoi\ge 1$. Black lines show trajectories of two particles (labelled ``1'' and ``2'') located in this plane. (bottom) Parallel and perpendicular components of the thermal velocity $\bb{w} \equiv \bb{v} - \bb{u}_{\rm i}$ measured with respect to the local magnetic field for each tracked particle versus time. The vertical dashed line marks the time of the snapshot, with the gray region indicating the time over which particle's trajectories are shown ($\approx$4 Larmor orbits). Plotted velocity tracks are filtered over 2 gyro-orbits to suppress fluctuations; unfiltered $w^2$'s, denoted by thin green lines, suggest potential fluctuations along the orbit.
\label{fig:track}
}
\end{figure*}

In this section we examine the turbulent heating of ions in our simulations. In doing so, we are careful to distinguish between {\em particle energization}, by which we mean the rate at which work is done on the particles by the electric field (viz., $Q\equiv\bb{v}\bcdot\bb{E}$) temporally averaged over long time intervals, and {\em particle heating}, by which we mean an increase in the Maxwellian temperature of the distribution function. This distinction is important. For example, while transit-time damping (TTD) is related to the work done by fluctuating perpendicular electric fields on Landau-resonant particles (i.e., perpendicular energization, $Q_\perp \equiv \bb{v}_\perp\bcdot\bb{E}_\perp$), it results in an increase in the {\em parallel} temperature $T_\parallel$. Likewise, changes in $T_\parallel$ need not be driven by particle energization; for example, such changes may occur via pitch-angle scattering of perpendicular particle energy into the parallel direction (as, indeed, is shown later in the paper).

Figure \ref{fig:heating_vs_k} shows time-averaged parallel ($Q_\parallel$) and perpendicular ($Q_\perp$) particle-energization rates as functions of $k_\perp\rhoi$. These are calculated by  using Fourier transforms to spatially filter the electric field into 12 logarithmically-spaced $k_\perp$-bins, giving $\bb{E}(k_{\perp,{\rm bin}})$, and then summing $v_\parallel E_\parallel(k_{\perp,{\rm bin}})$ and $\bb{v}_\perp\bcdot\bb{E}_\perp(k_{\perp,{\rm bin}})$ over all particles for each bin (the first and last bins are not shown).  (The use of Fourier transforms results in ``$\perp$" and ``$\parallel$" being measured here with respect to the guide field.) Rates are normalized to the rate of energy injection by the large-scale forcing, $Q_{\rm inj}$. In both runs, $Q_\perp \gg Q_\parallel$. For $\betaio = 1$, about 80\% of injected energy goes into the ions; the remaining 20\% is removed by hyper-resistivity. For $\beta_{\rm i0} = 0.3$, this ratio is roughly $75\%:25\%$. In both cases, $Q_\perp$ peaks at $k_\perp \rhoi\approx 4$, approximately where the measured spectral anisotropies in the KAW range imply $\omega_{\rm KAW} \approx \Omega_{\rm i0}$ (marked by the dotted lines in Figures \ref{fig:spectra}(g,h) and \ref{fig:heating_vs_k}).\footnote{These values were calculated using the approximation $\omega_{\rm KAW}=k_\parallel v_{\rm A} k_\perp\rho_{\rm i}/\sqrt{1+\beta_{\rm i}}$ and confirmed using the numerical linear hybrid-kinetic solver {\tt HYDROS} \citep[][cf.~their figure 8]{told16a}.} This suggests that the majority of the energization is produced when ions move in the oscillating potential of high-frequency KAWs, a possibility discussed further in \S\ref{sect:interpretation}. For now, we note that the sub-ion-Larmor spectrum of magnetic-field fluctuations seen in Fig.~\ref{fig:spectra}(a,b) steepens beyond the anticipated ${\approx}-2.8$ at the values of $k_\perp\rhoi$ for which the perpendicular energization is largest---likely not a coincidence. 

In Figure \ref{fig:track}, we present the velocity- and real-space tracks of two particles from the $\betaio = 1$ run, one (``1'') that exhibits appreciable perpendicular energization and another (``2'') that does not. Upper panels show snapshots of small-scale ($k_\perp \rhoi\ge 1$) magnetic (left) and electric (right) field fluctuations in the $x$-$y$ plane located at the $z$-coordinate of both tracked particles. The black lines trace projected particle trajectories over 4 gyro-periods. Particle 1 resides in a region of large-amplitude, sheet-like structures, whereas particle 2 samples only weak field fluctuations. (Note that the fluctuations experience the same $\bb{E}\btimes\bb{B}$ drift motion as the particles, and so particle 1's guiding center is not actually sweeping across the small-scale fluctuations.) Lower panels show the evolution of the particles' peculiar (``thermal'') velocity $\bb{w} \equiv \bb{v} - \bb{u}_{\rm i}$ over the final $1000\Omega_{\rm i0}^{-1}$ of the run. The gray shaded region marks the time interval over which particle trajectories are plotted. At each point, thermal velocities are averaged over $4\pi/\Omega_{\rm i0}$ to remove high-frequency oscillations in energy; these oscillations are shown for $w_{\rm tot}^2$ with the thin green line. 

Particle 1 sees an appreciable increase in its energy, with $\Delta w_{\rm tot}^2\approx 0.3 \vthio^2$ over only 4 gyro-periods. Most of this increase occurs perpendicularly to the local magnetic field. On the other hand, the energy of particle 2 stays nearly constant. Note further that the field-parallel and field-perpendicular energies do not grow monotonically. Often, they are subject to strong kicks during which the total energy is almost constant but the pitch angle of the particle changes dramatically, a feature we refer to as pitch-angle scattering. As a result, high-$w_\perp$ particles scatter and subsequently contribute to wings produced in the parallel distribution function, a feature we return to below in the context of Figures \ref{fig:dist}--\ref{fig:scatter}.

%
%
\begin{figure*}
\centering
\includegraphics[width=0.95\textwidth]{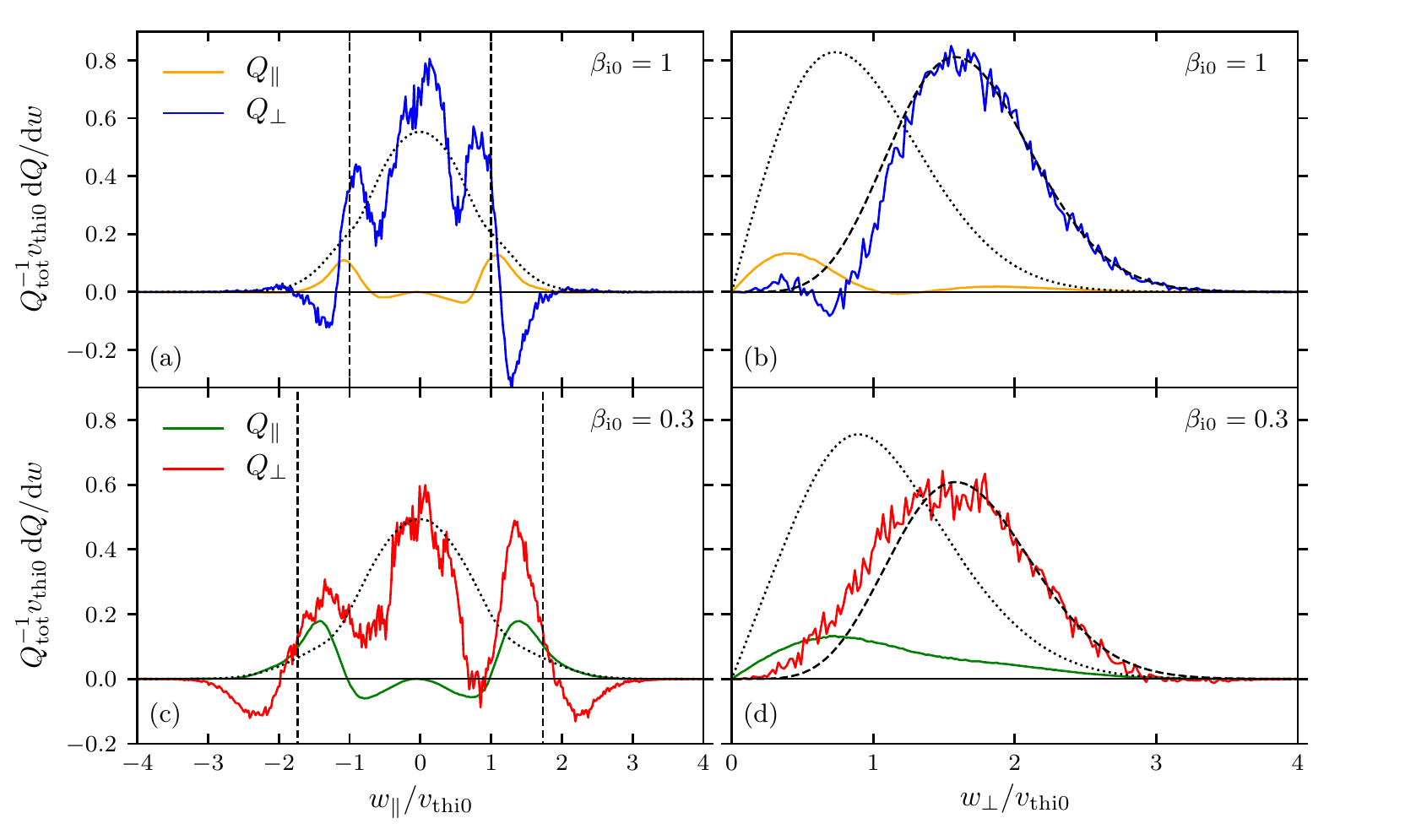}
\caption{Differential parallel ($\rmd Q_\parallel/\rmd w$) and perpendicular ($\rmd Q_\perp/\rmd w$) particle-energization rates as functions of parallel ($w_\parallel$) and perpendicular ($w_\perp$) particle thermal velocity for (a,b) $\betaio = 1$ and (c,d) $\betaio=0.3$. Rates are normalized to the total (parallel + perpendicular) energization rate. Dotted lines are $\langle f_{\rm i}(w_\parallel)\rangle$ and $\langle f_{\rm i}(w_\perp) \rangle$. Vertical dotted lines denote $\valfo$. The dashed lines in (b) and (c) are $(w_\perp/\vthio)^5 \exp(-w_\perp^2/\vthio^2)$, a reasonable fit to the data. In all panels, ``$\parallel$" and ``$\perp$" are measured with respect to the magnetic field at the location of each particle.
\label{fig:heating_vs_w}}
\end{figure*}
%
%
%

%
%
\begin{figure*}
\centering
\includegraphics[width=0.95\textwidth]{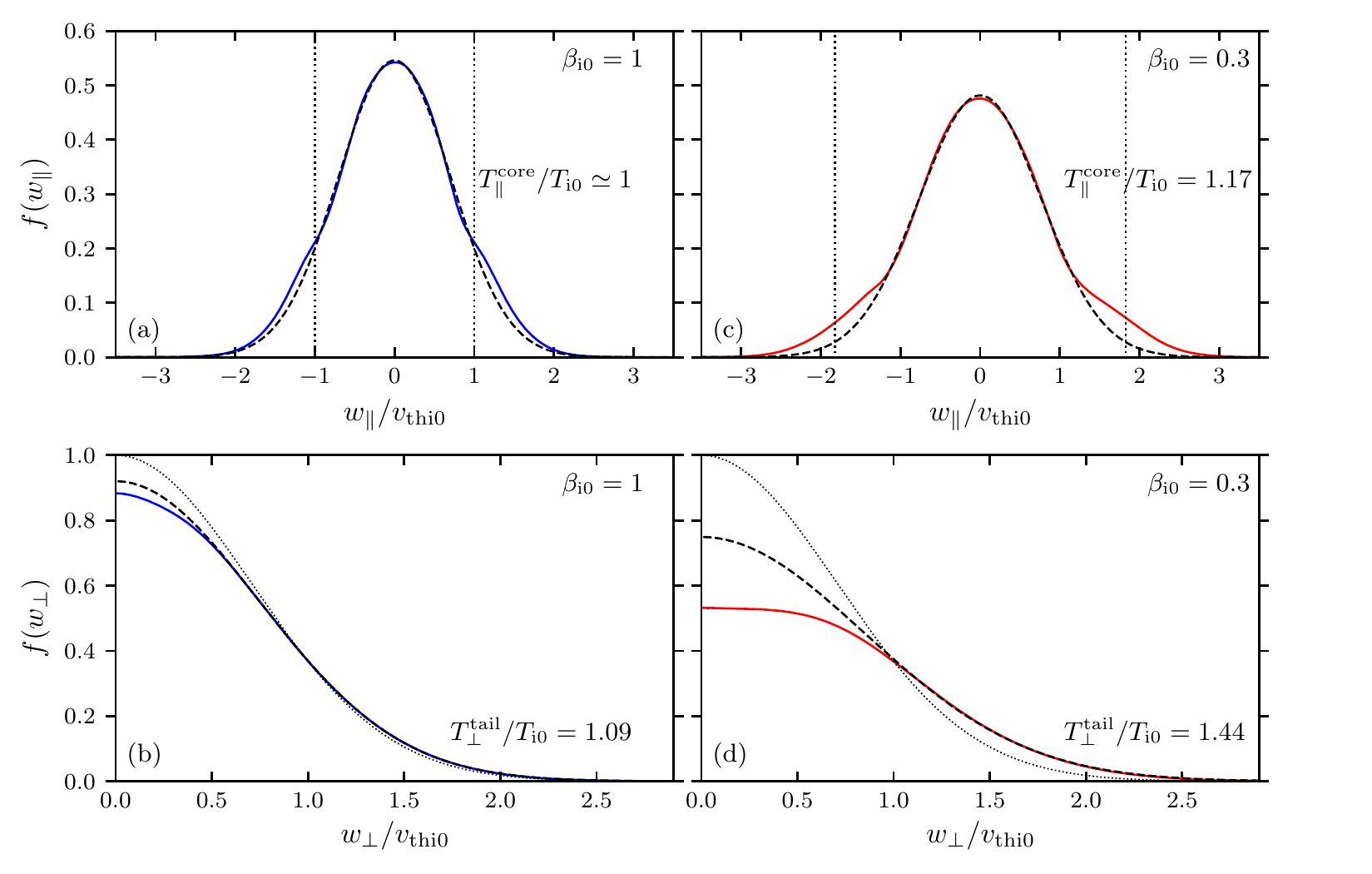}
\caption{(a,c) Box-averaged parallel distribution function $f(w_\parallel)$ at the end of $\betaio = 1$ and $0.3$ runs; both consist a Maxwellian core (dashed line, with fit temperature $T^{\rm core}_\parallel$) and non-Maxwellian wings. Vertical dashed lines denote $\valfo$. (b,d) Box-averaged perpendicular distribution function $f(w_\perp)$ for $\betaio=1$ and $0.3$; both consist of a Maxwellian tail (dashed line, with fit temperature $T^{\rm tail}_\perp$) and a flattened core. Dotted lines indicate the initial Maxwellians; ``$\parallel$" and ``$\perp$" are measured with respect to the  magnetic field at the location of each particle.
\label{fig:dist}}
\end{figure*}
%
%
%

%
%
\begin{figure}
\centering
\includegraphics[width=0.45\textwidth]{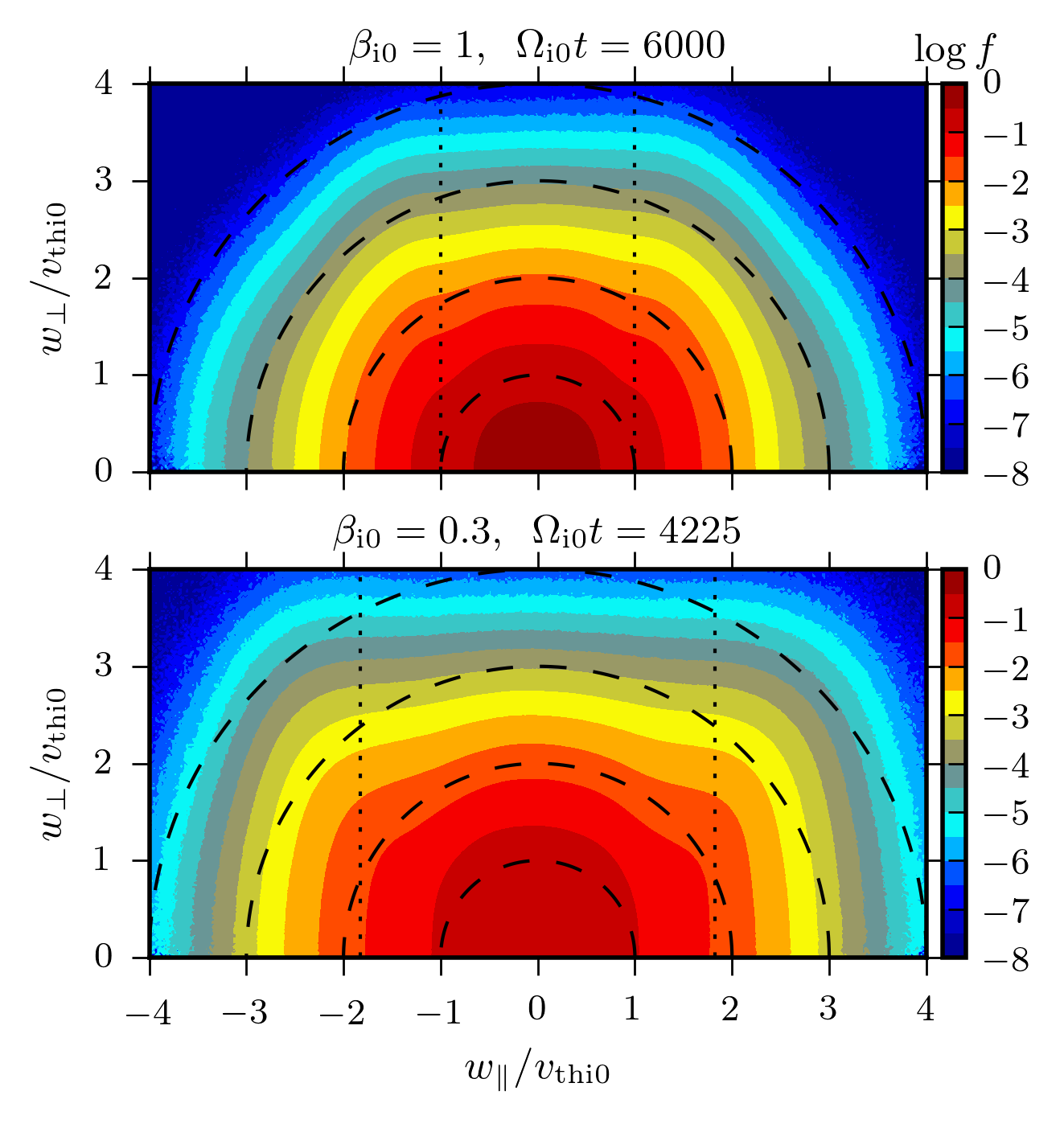}
\caption{Box-averaged 2D (gyrotropic) distribution function at the end of the $\betaio = 1$ and $0.3$ runs (cf.~Fig.~\ref{fig:dist}). Dashed lines trace constant-energy shells; the vertical dotted lines denote $w_\parallel = \valfo$.
\label{fig:dist2d}}
\end{figure}

Figure \ref{fig:heating_vs_w} shows the time-averaged parallel (perpendicular) differential energization, $\rmd Q_{\parallel (\perp)}/\rmd w$, in the gyrotropic velocity space $(w_\parallel,w_\perp)$. (Namely, $\rmd Q/\rmd w$ is the particle energization per interval of velocity space; these rates are normalized in the figure to the total energization rate $Q_{\rm tot}$ in each run. Here, ``$\parallel$" and ``$\perp$" are measured with respect to the {\em local} magnetic field at the position of the particle.) Figure \ref{fig:heating_vs_w} is to be read alongside Figure \ref{fig:dist}, which displays the parallel and perpendicular ion distribution functions measured at the ends of both runs, {\it viz.}, $f(w_\parallel) \equiv \int\rmd w^2_\perp \,f(w_\parallel,w_\perp)$ and $f(w_\perp) \equiv \int\rmd w_\parallel \, f(w_\parallel,w_\perp)$, respectively (again, measured with respect to the local magnetic-field direction). The dashed lines in Figure \ref{fig:dist} show best-fit Maxwellians to the core of $f(w_\parallel)$ (with temperature $T_\parallel^{\rm core}$) and to the tail of $f(w_\perp)$ (with temperature $T_\perp^{\rm tail}$). Both runs exhibit the following attributes: (i) resonant features in the particle-energization rates with fine-scale structure near $w_\parallel \sim \vthio$ and ${\sim}\valfo$ (Figs \ref{fig:heating_vs_w}(a),(c)); (ii) quasi-linear flattening of the parallel distribution at $|w_\parallel|/\valfo \sim 1$ and non-thermal wings at $|w_\parallel|/\vthio \gtrsim 1$ (Figs \ref{fig:dist}(a),(c)); (iii) almost no change in the parallel temperature of the core ($|w_\parallel|/\vthio < 1$; Figs \ref{fig:dist}(a),(c)), with very little parallel energization there (Figs \ref{fig:heating_vs_w}(a),(c)); (iv) a broadened perpendicular distribution for $1 \lesssim w_\perp/\vthio \lesssim 3$ (Figs \ref{fig:dist}(b),(d)), where the perpendicular energization peaks (Figs \ref{fig:heating_vs_w}(b),(d)); and (v) a flattening of the core of the perpendicular distribution $f(w_\perp)$ (Figs \ref{fig:dist}(b),(d)), with suppressed perpendicular energization at $w_\perp/\vthio \lesssim 1$ (Figs \ref{fig:heating_vs_w}(b),(d)). Regarding this final point, we find evidence early in each run for perpendicular energization at $w_\perp/\vthio \lesssim 1$, which ultimately causes the observed flattening the core of $f(w_\perp)$ (see Fig.~\ref{fig:diffusion} and accompanying discussion below). 

All of these features can also be seen in the full 2D (gyrotropic) distribution function $f(w_\parallel,w_\perp)$ shown in Figure \ref{fig:dist2d}, where the differences between the two runs are even more striking. In particular, non-Maxwellian features such as the flattened $w_\perp$ core and the parallel beams at $w_\parallel \sim \valfo$ are readily apparent, with the latter driving $\partial f/\partial w_\parallel \approx 0$ for $w_\perp/\vthio \gtrsim 3$ and $|w_\parallel|\lesssim \valfo$.

We associate with these non-Maxwellian features a number of physical effects. The early-time flattening of $f(w_\perp)$ for $w_\perp/\vthio \lesssim 1$ is most likely due to stochastic heating by low-frequency, Larmor-scale fluctuations, following the prediction by \citet{klein16} that such heating leads to a flattop distribution (see also \citealt{jc01}). As stochastic heating is expected to be more important at lower $\beta$ for fixed $\delta B_\perp/B_0$ \citep[e.g.,][]{chandran10,hoppock2018}, it is noteworthy that $f(w_\perp)$ exhibits a larger amount of flattening for $\betaio=0.3$ than for $\betaio=1$. The broadening of the thermal tail of $f(w_\perp)$ is instead driven by the peak in $\rmd Q_\perp/\rmd w_\perp$ at $w_\perp/\vthio \approx 1.6$, which constitutes the bulk of the perpendicular heating and is related to the peak in $\rmd Q/\rmd\log k_\perp$ at $k_\perp\rhoi\approx 4$--$5$ (Fig.~\ref{fig:heating_vs_k}).

%
%
\begin{figure*}
\centering
\includegraphics[width=0.95\textwidth]{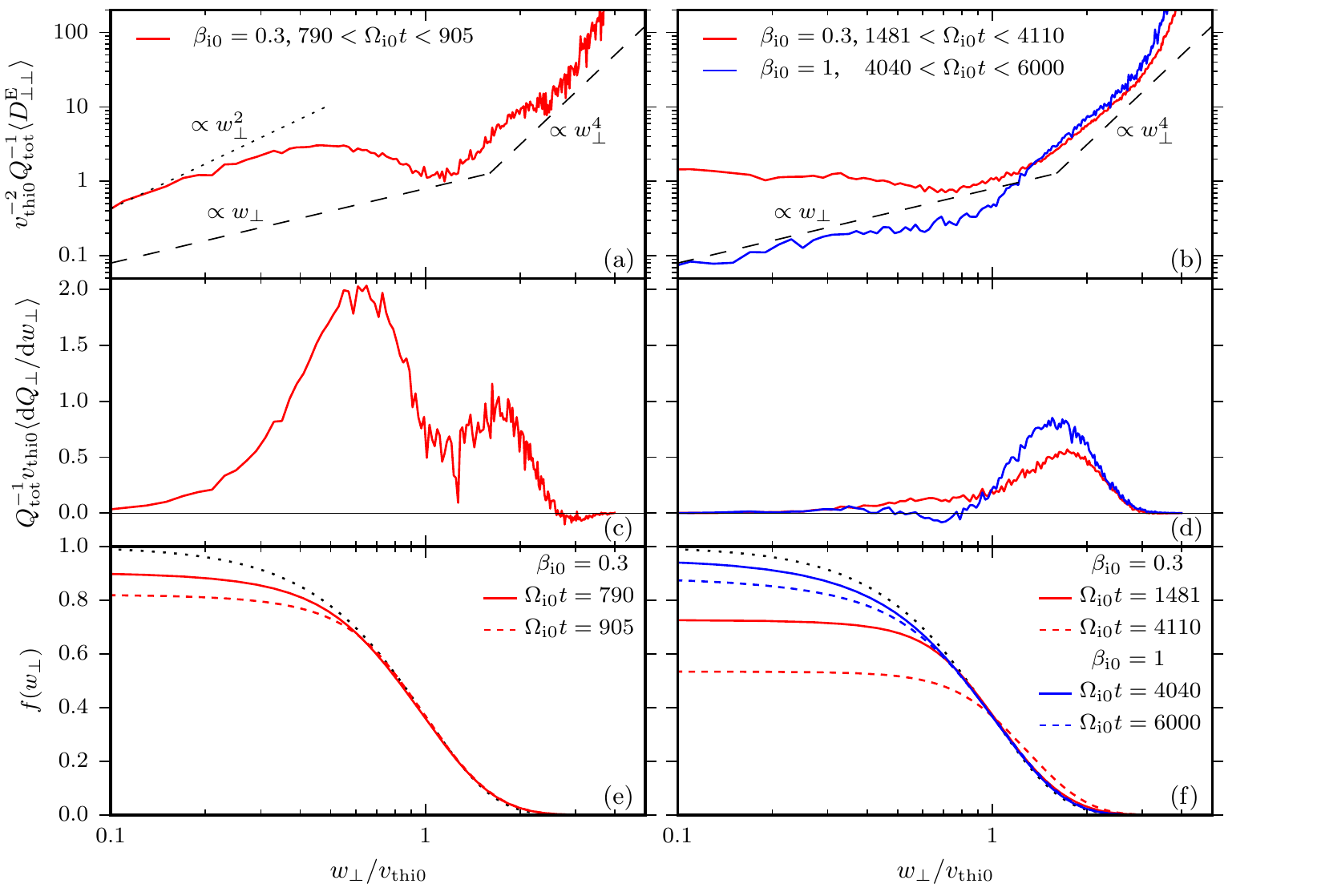}
\caption{Box-averaged perpendicular-energy diffusion coefficient, $\langle D_{\perp\perp}^{\rm E}\rangle$ (see Equation \ref{eqn:diff}), averaged over two time windows: (a) at early times, during which the core of the perpendicular distribution function becomes appreciably flattened, and (b) at late times after $f(w_\perp)$ is cored and during which its temperature steadily grows (the stated values of $T^{\rm tail}_\perp/T_{\rm i0}$ are obtained from a Maxwellian fit to the $w_\perp/\vthio >1$ tail of the distribution function). At late times, the diffusion coefficient is flat for $w_\perp < \vthio$; because ${\rm d}f(w_\perp)/{\rm d}w_\perp \sim 0$ in this range, very little heating happens at small velocities. At larger velocities, $\langle D_{\perp\perp}^{\rm E}\rangle\propto w_\perp^4$ seems to be a fair approximation.  In all panels, ``$\perp$" is measured with respect to the magnetic field at the location of each particle.
\label{fig:diffusion}}
\end{figure*}

This interpretation is further strengthened by examining the perpendicular-energy diffusion coefficient $D^{\rm E}_{\perp\perp}$. Assuming that the distribution function evolves according to a Fokker--Planck-like equation,
\begin{equation}
    \frac{\partial f}{\partial t} = \frac{\partial}{\partial e_\perp} \left(D^{\rm E}_{\perp\perp} \frac{\partial f}{\partial e_\perp} \right),
\end{equation}
where $e_\perp \equiv  w_\perp^2/2$ is the field-perpendicular kinetic energy per particle, the energization rate
\begin{align}
    Q_\perp &\equiv \frac{\partial}{\partial t} \int\rmd e_\perp \, e_\perp f(e_\perp)  =\int\rmd e_\perp \,  e_\perp \frac{\partial}{\partial e_\perp} \left(D^{\rm E}_{\perp\perp} \frac{\partial f}{\partial e_\perp} \right) \nonumber\\
     &=- \int\rmd e_\perp \, D^{\rm E}_{\perp\perp} \frac{\partial f}{\partial e_\perp} ,
\end{align}
where the final equality is obtained after integrating by parts. The box-averaged perpendicular-energy diffusion coefficient can thus be computed from the derivatives of the perpendicular energization and the distribution function, as follows:

\begin{equation}\label{eqn:diff}
    \langle D^{\rm E}_{\perp\perp}\rangle \equiv - \left\langle\pD{e_\perp}{Q_\perp}\right\rangle \bigg/ \left\langle \pD{e_\perp}{f}\right\rangle .
\end{equation}
This quantity is plotted as a function of $w_\perp/\vthio$ in Figure \ref{fig:diffusion} both at early times in the $\betaio=0.3$ run (panel a) and at late times in both runs (panel b). Accompanying these panels are (c,d) the differential perpendicular particle energization (as in Figure \ref{fig:heating_vs_w}) and (e,f) the perpendicular distribution function (as in Figure \ref{fig:dist}) at those times. In the latter panels, the evolution of $f(w_\perp)$ from its initial condition (black dotted line) to its profile at the beginning (solid line) and the end (dashed line) of the time-averaging window are shown. At early times, the diffusion coefficient and the accompanying $\rmd Q_\perp/\rmd w_\perp$ exhibit two distinct peaks. We associate the peak at $w_\perp/\vthio < 1$ with low-frequency stochastic heating, which flattens the core of $f(w_\perp)$ by accelerating particles to larger $w_\perp$. The $w_\perp/\vthio > 1$ peak in $\rmd Q_\perp/\rmd w_\perp$ is associated with a diffusion coefficient that scales approximately as $w^4_\perp$. At late times in the $\betaio=0.3$ run (red lines), after the core of the perpendicular distribution has been flattened, the diffusion coefficient is roughly constant with $w_\perp$ and very little energization happens for $w_\perp/\vthio < 1$ (note that $\langle D^{\rm E}_{\perp\perp}\rangle$ is not well-defined in this limit, as both $\partial Q_\perp/\partial w^2_\perp$ and $\partial f/\partial w^2_\perp$ are close to zero). At larger velocities, the perpendicular energization peaks (as in Figure \ref{fig:heating_vs_w}) and the diffusion coefficient continues to scale approximately as $w^4_\perp$. In the $\betaio=1$ run, which did not experience as great of a flattening in the core of $f(w_\perp)$, $\langle D^{\rm E}_{\perp\perp}\rangle \propto w_\perp$ appears to be a good approximation for $w_\perp/\vthio<1$. 

Most of the injected energy goes into the increase of perpendicular temperature and into the development of non-thermal tails in the parallel distribution function. (Overall, ${\approx}60\%$ of the total energization $Q_{\rm tot}$ ultimately finds its way into the non-thermal $v_\parallel$ wings, with the remaining ${\approx}40\%$ going into raising the perpendicular temperature.) Given that most of the energization is perpendicular, this strongly suggests that transit-time damping (TTD) and pitch-angle scattering of super-thermal $w^2_\perp$ into $w^2_\parallel$ are the mechanisms responsible for the non-Maxwellian features seen in $f(w_\parallel)$.

We tested these possibilities by examining the energization and pitch angles of 160,000 individually tracked particles in the $\betaio=1$ run. We found that the mirror force $\mu\,\rmd B(\bb{x}_{\rm p})/\rmd t$ (where $\mu\equiv w^2_\perp/2B$ is the magnetic moment) is responsible for ${\lesssim}20\%$ of $Q_\perp$, a value consistent with the quasi-linear flattening of the  distribution function observed at $w_\parallel \sim \valfo$ in Figures \ref{fig:dist} and \ref{fig:dist2d}. The remaining (small) amount of increase in $T_\parallel$ can be accounted for by Landau damping (i.e., $Q_\parallel = v_\parallel E_\parallel \lesssim 0.1 Q_{\rm tot}$). The other ${\gtrsim}80\%$ of $Q_\perp$ is due to some other mechanism that leads to perpendicular heating at $w_\perp > \vthio$ (see \S\ref{sect:interpretation} for a discussion of possibilities). To examine the effect of pitch-angle scattering on the distribution function, we divided each particle track into time intervals of $5\cdot 2\pi/\Omega_{\rm i0}$ and computed $w^2_\perp/w^2_\parallel$ at the beginning and the end of each interval. We then filtered each interval based on whether or not $w^2_\perp/w^2_\parallel$ changes (up or down) by a certain minimal factor that we call ``threshold''. Large values of threshold represent large changes in a particle's pitch angle; small values of threshold include time intervals during which the pitch angle is almost constant. Figure \ref{fig:scatter} shows $\overline{\Delta w^2_{(\parallel,\perp,{\rm tot})}/\Delta t}$, the average\footnote{To compute $\overline{\Delta w_{(\parallel,\perp)}^2/\Delta t}$ as a function of threshold, we sum $\Delta w_\parallel^2$ and $\Delta w_\perp^2$ for all events that change $w^2_\perp/w^2_\parallel$ by a factor larger than threshold, and then divide this energy increment by the total time over which we examine the particle tracks ($\Delta t = 2000\Omega_{\rm i0}^{-1}$) and by the total number of tracked particles.\label{ftnote:scatter}} rate of change of the parallel (blue), perpendicular (red), and total (green) thermal energies of tracked particles, segregated into super- and sub-thermal populations ($w_\perp>\vthio$ and $w_\perp<\vthio$, respectively). Given a threshold value, Figure \ref{fig:scatter} provides the expected energization rates for an individual particle due to all events whose change in $w^2_\perp/w^2_\parallel$ are above that threshold. As the length of the time interval used to sub-divide the particle tracks is somewhat arbitrary, we indicate with the shaded regions the variation of particle-energization rates for time intervals from $4\cdot 2\pi/\Omega_{\rm i0}$ to $8\cdot 2\pi/\Omega_{\rm i0}$. Note that $|\overline{\Delta w^2_{\parallel,\perp}/\Delta t}|$ is a decreasing function of threshold, as it represents the cumulative energization for all events above the threshold. The energization per event (not shown) is an increasing function of threshold. For $w_\perp<\vthio$, there is a net conversion of thermal energy from parallel to perpendicular, consistent with the other diagnostics shown in Figures \ref{fig:heating_vs_w}--\ref{fig:diffusion}. For $w_\perp>\vthio$, however, the flow of thermal energy between parallel and perpendicular is reversed and, for thresholds ${\gtrsim}1.5$, this flow takes place {\em at constant total energy}. This strongly suggests that pitch-angle scattering is responsible for this transfer of super-thermal perpendicular energy into super-thermal parallel energy \citep[similar to recent work by][]{isenberg2019}, ultimately producing the non-thermal wings seen in Figures \ref{fig:dist} and \ref{fig:dist2d}.\footnote{Evidence of pitch-angle scattering can also be seen in the evolution of particle 1 at $\Omega_{\rm i0} t \approx 5400$--$5600$ in Figure \ref{fig:track}; and there is the additional circumstantial evidence that the more pronounced non-thermal wings in $f(w_\parallel)$ seen in the $\betaio=0.3$ run (as compared to the $\betaio=1$ run) coincides with a larger perpendicular temperature measured in the tail of $f(w_\perp)$.}

%
%
\begin{figure}
\centering
\includegraphics[width=0.45\textwidth]{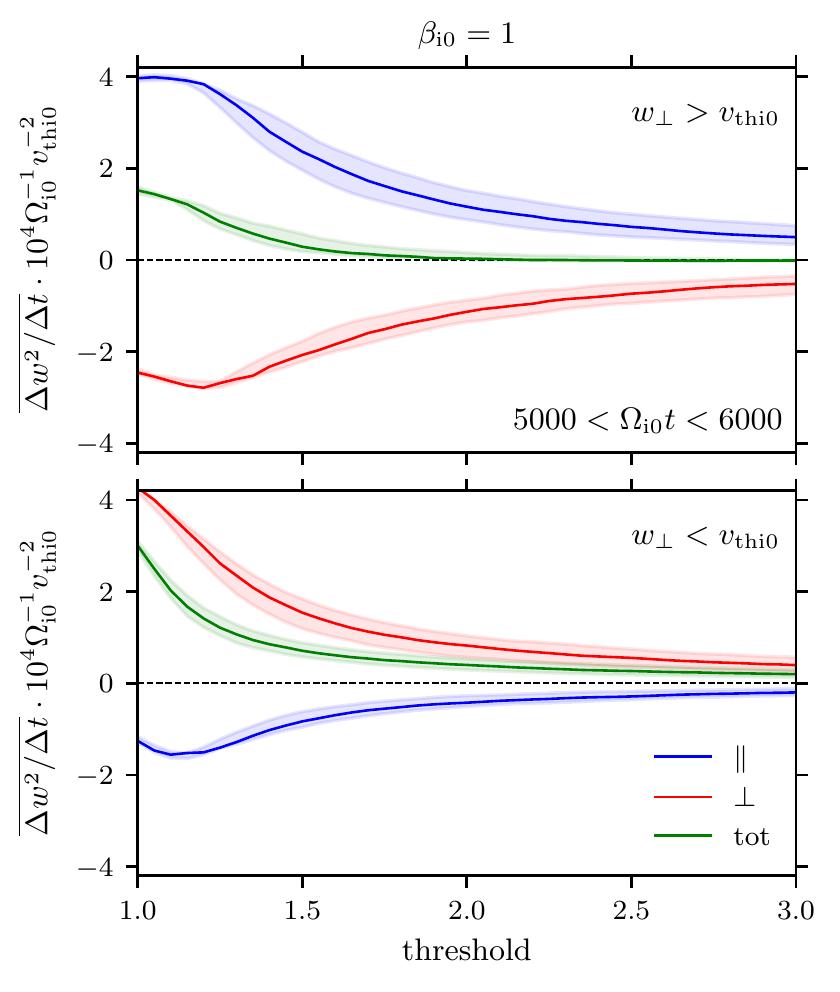}
\caption{
The rate of change of parallel (blue), perpendicular (red), and total (green) thermal energy of 160,000 tracked particles from the $\betaio = 1$ run, segregated into super-thermal ($w_\perp>\vthio$; top) and sub-thermal ($w_\perp<\vthio$; bottom) populations resulting from all events above ``threshold'', the minimum factor by which $w^2_\perp/w^2_\parallel$ changes (up or down) within a time interval $5\cdot 2\pi/\Omega_{\rm i0}$ (see footnote \ref{ftnote:scatter}). The shaded regions indicate the variation of these rates as a function of time interval (from $4\cdot 2\pi/\Omega_{\rm i0}$ to $8\cdot 2\pi/\Omega_{\rm i0}$). For $w_\perp > v_{\rm thi0}$ in particular, $w_\parallel^2$ and $w_\perp^2$ change much more than does $w_{\rm tot}^2$, consistent with pitch-angle scattering. (Here, ``$\parallel$" and ``$\perp$" are measured with respect to the magnetic field at the location of each particle.)
\label{fig:scatter}}
\end{figure}

\begin{figure}
\centering
\includegraphics[width=0.45\textwidth]{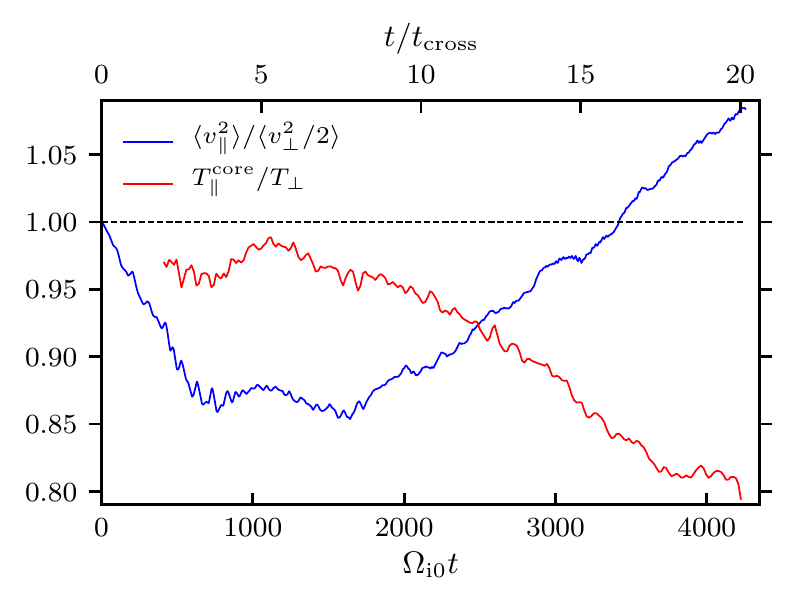}
\caption{Time evolution from the $\betaio=0.3$ run of the ratios of (blue line) parallel-to-perpendicular particle energy, $\langle v_\parallel^2\rangle/\langle v_\perp^2/2\rangle$, and (red line) parallel-to-perpendicular best-fit bi-Maxwellian temperatures, $T^{\rm core}_\parallel/T_\perp$. In the latter, $T^{\rm core}_\parallel$ describes only the thermal core of the parallel distribution function, excluding the non-thermal wings. (Recall that $\bb{v}$ includes both the bulk and thermal velocities of the ion particles; ``$\parallel$" and ``$\perp$" are measured with respect to the magnetic field at the location of each particle.)
\label{fig:energies}}
\end{figure}

We re-emphasize that, while $T_\perp > T_\parallel^{\rm core}$ at the end of our simulations, more than half of the cascade energy ultimately goes into the development of the non-thermal wings in the parallel distribution function. To illustrate that, the blue line in Figure \ref{fig:energies} traces the temporal evolution of the ratio of the total parallel and perpendicular energies of the particles, $\langle v_\parallel^2\rangle/\langle v_\perp^2/2\rangle$, from the $\betaio=0.3$ run in which the non-thermal wings are most pronounced (the factor of $1/2$ accounts for the number of degrees of freedom in the perpendicular direction, so that an isotropic Maxwellian has $\langle v_\parallel^2\rangle/\langle v_\perp^2/2\rangle = 1$). The accompanying red line represents the ratio of best-fit Maxwellian temperatures to the core of the parallel distribution function ($T_\parallel^{\rm core}$) and to the tail of the perpendicular distribution function ($T_\perp^{\rm tail}\simeq T_\perp$). The initial drop in $\langle v_\parallel^2\rangle/\langle v_\perp^2/2\rangle$ is caused by the increase in perpendicular bulk motion in the initial stage of the simulation. (Recall that $\bb{v}$ denotes the total (bulk + thermal) velocity of the ion particles.) Once the turbulence obtains quasi-steady state (for $t/t_{\rm cross}\gtrsim 3$), the perpendicular temperature steadily grows larger than the parallel temperature of the core. This is mirrored in the evolution of $\langle v^2_\parallel \rangle / \langle v^2_\perp/2 \rangle$, at least until $t/t_{\rm cross} \approx 8$, after which the ratio of energies suddenly begins to increase, eventually becoming larger than 1 at $t/t_{\rm cross} \approx 16$. The distinction between the ratio of energies and the ratio of best-fit-Maxwellian temperatures is thus an important one. Indeed, while non-thermal $v_\parallel$-wings are often observed in ion velocity distribution functions in the solar wind, the ``observed'' $T_\parallel$ is often determined by a bi-Maxwellian fit to the core of the distribution \citep[e.g.,][]{bame75,marsch1982}.

%
%
%
%
%


\section{Interpretation of Sub-Ion-Larmor-Scale Perpendicular  Heating}\label{sect:interpretation}

The perpendicular energization that occurs at early times for $w_\perp \lesssim \vthio$ appears to be consistent with the  predictions of stochastic ion heating by low-frequency fluctuations. Namely, the core of the perpendicular distribution function becomes flattened as particles there are promoted to higher perpendicular energies via $\bb{v}_\perp\bcdot\bb{E}_\perp$ energization, with lower $\betaio$ and larger amplitudes leading to more flattening. This finding is notable, in that it was achieved in a self-consistent setting in which the evolution of the distribution function is allowed to feed back on the electromagnetic fields that drive that evolution. This is in contrast to other studies of stochastic heating that have employed a test-particle approach.

By contrast, the origin of the perpendicular energization that occurs for $w_\perp \gtrsim \vthio$ is less clear. Despite the highly suggestive correlation between the $k_\perp\rhoi$ at which (i) the perpendicular ion heating peaks, (ii) the sub-ion-Larmor magnetic-energy spectrum steepens beyond the canonical $-2.8$, and (iii) the linear KAW frequency attains the ion-Larmor frequency, it is nevertheless difficult for us to definitively conclude that the majority of the ion heating is due to cyclotron heating by high-frequency KAW turbulence. The reason for this reluctance is twofold. First, while the scale separation and dynamic range afforded by our simulations bears resemblance to solar-wind data on the $k_\perp \rhoi \lesssim 1$ side, on the $k_\perp \rhoi \gg 1$ side there is a less-than-optimal scale separation between the grid scale (near which hyper-resistivity and low-pass filtering are important) and the location of maximal perpendicular energization and the  concomitant spectral steepening. But more important than this computational concern is a physical one: we currently have no theory explaining the dependencies of the perpendicular particle-energization rate and the perpendicular energy-diffusion coefficient on the particles' perpendicular energy for $w_\perp \gtrsim \vthio$. Both of these profiles appear to be inconsistent with previous studies of stochastic heating \citep{klein16} and cyclotron damping \citep[e.g.,][]{isenberg04,isenbergvasquez07,isenbergvasquez11,isenbergvasquez15}, for which $D^{\rm E}_{\perp\perp} \propto w_\perp^2$.

That being said, it is notable that {\it Stereo} measurements in high-speed solar-wind streams show a rapid decrease in the power anisotropy (i.e., the difference between the energy stored in the perpendicular and parallel fluctuations at a given scale) near $2~{\rm Hz}$, which \citet{podesta09} associated with the strong linear dissipation of KAWs occurring at $k_\perp\rhoi\approx 4$ when $k_\perp/k_\parallel=10$ (parameters not unlike ours). It is in this wavenumber range that KAWs are known to couple to ion-Bernstein waves (IBWs; \citealt{bernstein58}), which \citet{podesta12} showed provide a channel for mode coupling and thus energy transfer from KAWs. Once excited, IBWs are strongly damped through a combination of ion-cyclotron and electron-Landau resonances (the latter of which are of course absent in our simulations). This is rather suggestive, but more work is needed to predict the velocity-space signatures of ion energization by high-frequency KAWs/IBWs.

\section{Summary: Implications for the Solar Wind and for Simulations of Kinetic Turbulence}  
\label{sect:summary}

We have presented 3D numerical simulations of driven, quasi-steady-state, hybrid-kinetic turbulence in a magnetized plasma of relevance to the $\beta_{\rm i} \lesssim 1$ solar wind. Despite the more general hybrid-kinetic framework employed, many aspects of the spectral scalings are in rough agreement with those obtained using gyrokinetics \citep{howes08,howes11,told15}. These include a critically balanced, spatially anisotropic, inertial-range cascade of Alfv\'{e}nic fluctuations; a spectral break occurring near the ion-Larmor scale, beyond which the magnetic spectrum steepens and the electric spectrum flattens; signatures of linear phase mixing in the ion distribution function (e.g., flattening of the parallel distribution function near $v_\parallel \sim \valf$); and what appears to be a sub-ion-Larmor-scale cascade composed primarily of KAWs. However, there are differences, most notably in the efficiency and mechanism of ion heating.

Although these simulations were originally designed to test the theory of stochastic ion heating occurring at $k_\perp \rho_{\rm i} \sim 1$---which we do find---our results also make the case for perpendicular ion heating {\em at sub-ion-Larmor scales}, as a cascade of KAWs approaches the ion-cyclotron frequency. This heating, alongside contributions from TTD, Landau damping, and pitch-angle scattering, simultaneously heat the particles perpendicularly, produce non-Maxwellian wings in the ion parallel distribution function, and steepen the sub-ion-Larmor-range magnetic-energy spectra beyond the typically observed $-2.8$ power-law scaling by transferring electromagnetic energy into thermal energy. It is perhaps no coincidence, then, that the ion-kinetic-range spectral index as measured in the solar wind correlates with the amount of inferred energy dissipation, with more dissipation going hand-in-hand with steeper spectra \citep[e.g.,][]{smith06}. We predict that such spectra will be found to be accompanied by non-Maxwellian wings in $f(w_\parallel)$, a preponderance of perpendicular heating (over parallel heating), and a flattened core in $f(w_\perp)$.

Given that complementary gyrokinetic theory and simulations show a predominance of electron heating over ion heating for $\beta \lesssim 1$ \citep{howes10,howes11,told15,navarro16,kawazura18}, it is worthwhile to contemplate which framework is better suited for describing near-Earth solar-wind turbulence (see \citet[][\S 3]{howes08heating} for arguments in favor of a gyrokinetic description). Gyrokinetic theory is built on the assumption of asymptotically low-frequency, small-amplitude, spatially anisotropic fluctuations; the accompanying computational savings affords a magneto-kinetic description of a realistic hydrogenic plasma without prohibitive cost. However, the magnetic moment is an invariant in the gyrokinetic equations (in the absence of explicit collisions), and so perpendicular particle heating is not allowed. By contrast, the hybrid model makes no such simplifying assumptions, but saves computational expense by neglecting electron kinetics. While the latter precludes a truly rigorous study of ion versus electron heating, it is worth noting that there is empirical evidence for a majority fraction ($\gtrsim$60\%) of ion versus electron heating between $0.3$ and $5~{\rm au}$ in the solar wind \citep{cranmer09}; recall that ${\approx}75$--$80\%$ of our cascade energy goes into heating the ions (the rest is dissipated via hyper-resistivity). Moreover, the ion-Larmor-scale spectral anisotropy in the $\beta_{\rm i} \sim 1$ solar wind is not necessarily asymptotically small. Indeed, $\theta_{(\bs{k},\bs{B})} \approx 80^\circ$--$90^\circ$ for $f_{\rm spacecraft} \sim 1~{\rm Hz}$ fluctuations measured in the near-Earth solar wind \citep{sahraoui10}; our values are similar, $\theta_{(\bs{k},\bs{B})} \approx 85^\circ$--$86^\circ$. If $k_\parallel$ scales with $k_\perp$ to some power larger than the $1/3$ adopted by \citet{howes08heating} for $k_\perp \rho_{\rm i0} > 1$, as it does in our simulations and as is predicted in theories of intermittent KAW turbulence \citep{bp12}, it becomes all the more probable that the ion-cyclotron frequency will be attained in a KAW cascade. These considerations favor the hybrid-kinetic description over the gyrokinetic one. 

More broadly, our study of ion heating in kinetic, Alfv\'{e}nic turbulence may be applicable to a number of collisionless astrophysical plasmas in which the collisional mean free path is comparable to or even larger than the system size, the canonical example of which being the low-luminosity accretion flow onto the supermassive black hole at the Galactic center, Sgr A$^\ast$. The observed low luminosity of this system can be explained if the gravitational energy released during accretion is stored primarily in the poorly radiating ions rather than the electrons \citep{ichimaru1977,narayanyi1994,narayan1998}. As angular-momentum transport in myriad accretion disks is thought to be driven by magnetorotational turbulence \citep{bh98}, the question of ion versus electron heating in such turbulent flows is thus important for models of low-luminosity accretion \citep{qg99,sharma07}. Recently, there have been several studies implementing various particle-heating prescriptions in general relativistic MHD simulations of black-hole accretion flows \citep[e.g.,][]{ressler2015,sadowski2017,chael2018}. These ``sub-grid'' prescriptions are based either on gyrokinetic theory and simulation \citep{howes10,told15}, which predict preferential electron (ion) heating at low (high) plasma beta driven by Landau-resonant damping, or on models of collisionless reconnection \citep{rowan2017,werner2018,rowan2019}, in which the amount of ion versus electron heating is sensitive to the presence of a strong guide field. It would be interesting to study the effect of the non-Landau-resonant processes presented in this paper, which predominantly heat the ions, on the imaging and evolution of collisionless accretion flows. However, the application of our results to these systems is not straightforward. While the wavevector anisotropies in our simulations are consistent with the scale separation observed in the solar wind ({\it viz.}, a factor of ${\sim}10^4$ between the outer scale and the ion-Larmor scale), the scale separation in black-hole accretion flows is expected to be even larger (e.g., a factor of ${\sim}10^7$ between the outer scale and the ion-Larmor scale for Sgr A$^\ast$; see \citealt{quataert1998}). This implies that Alfv\'{e}n-wave/KAW frequencies near the ion-Larmor scale are a factor ${\sim}10$ smaller than in the solar wind, possibly inhibiting particle heating via cyclotron resonances. We defer a study of ion heating in this regime to future work.

In a subsequent publication, a wider parameter study will be conducted alongside further analysis of field-particle correlations (following \citealt{kleinhowes16}, \citealt{howes17}, and \citealt{klein17}). In the meantime, we hope that the various particle energization diagnostics employed herein will spur their application to both existing and future simulation data of solar-wind-like kinetic turbulence.

\smallskip
\acknowledgements

L.A.~and M.W.K.~were supported by the National Aeronautics and Space Administration (NASA) under Grant No.~NNX16AK09G issued through the Heliophysics Supporting Research Program. Additional support for M.W.K.~was provided by an Alfred P.~Sloan Research Fellowship in Physics and, during the early stages of this project (c.~2014), by a Lyman Spitzer, Jr.~Fellowship. B.D.G.C.~was supported by NASA grants NNN06AA01C, NNX15AI80, NNX16AG81G, and NNX17AI18G, and by the National Science Foundation (NSF) under grant PHY-1500041. E.Q.~was supported in part by a Simons Investigator award from the Simons Foundation. High-performance computing resources were provided by: the Texas Advanced Computer Center at The University of Texas at Austin under grant numbers TG-AST140011 and TG-AST130058; the NASA High-End Computing (HEC) Program through the NASA Advanced Supercomputing (NAS) Division at Ames Research Center; and the PICSciE-OIT TIGRESS High Performance Computing Center and Visualization Laboratory at Princeton University. This work used the Extreme Science and Engineering Discovery Environment (XSEDE), which is supported by NSF grant OCI-1053575. This work benefited from useful conversations with Christopher Chen, Alfred Mallet, Robert Wicks, Vladimir Zhdankin, and especially Silvio Sergio Cerri, Gregory Howes, Philip Isenberg, Kristopher Klein, and Alexander Schekochihin.


\bibliographystyle{apj}
\bibliography{references}

\end{document}